\documentclass{amsart}
\usepackage{amsfonts,amssymb,amsmath,amsthm,mathrsfs}
\newtheorem{thrm}{Theorem}[section]

\newtheorem{prop}[thrm]{Proposition}

\theoremstyle{definition}

\begin{document}
\author[C.~A.~Mantica and L.~G.~Molinari]
{Carlo~Alberto~Mantica and Luca~Guido~Molinari}
\address{C.~A.~Mantica: 
I.I.S. Lagrange, Via L. Modignani 65, 20161 Milano, 
and I.N.F.N. sez. Milano, Italy -- L.~G.~Molinari: Physics Department `Aldo Pontremoli',
Universit\`a degli Studi di Milano and I.N.F.N. sez. Milano,
Via Celoria 16, 20133 Milano, Italy.}
\email{carlo.mantica@mi.infn.it, luca.molinari@unimi.it}
\subjclass[2010]{Primary 83C20, Secondary 53B30}
\keywords{Twisted spacetime, warped spacetime, torqued vector}

\begin{abstract}
In this note we characterize 1+n doubly twisted spacetimes in terms of `doubly torqued' vector fields. They extend 
Bang-Yen Chen's characterization of twisted and generalized Robertson-Walker spacetimes with torqued and concircular vector fields. The result is a simple classification of 1+n doubly-twisted, doubly-warped, twisted and generalized Robertson-Walker spacetimes.
\end{abstract}
\title[A simple characterization of doubly twisted spacetimes]{A simple characterization\\ of doubly twisted spacetimes}
\maketitle

\section{Introduction}
Several interesting Lorentzian metrics have a block-diagonal form, with time labelling a foliation with spacelike hypersurfaces. They include doubly-twisted, doubly-warped, twisted, warped, and Robertson-Walker spacetimes \cite{Ponge}. The Frobenius theorem characterizes the vector fields $u_i$ that are hypersurface orthogonal, for which there exist functions $\lambda $ and $f$ such that, locally, $\lambda u_i =\nabla_i f$ (\cite{Stephani}, p.19). This establishes a dual description of such spacetimes: the special form of the metric allows explicit evaluations, the one in terms of the vector field is covariant. While physicists conceive the vector field as a congruence of timelike trajectories, 
geometers prefer other vectors, as is here illustrated.

Doubly twisted spacetimes were introduced (and named `conformally separable') by Kentaro Yano in 1940: 
\begin{align}
ds^2 = -b^2({\bf q},t) dt^2 +a^2(t,{\bf q})g^*_{\mu\nu}({\bf q})dq^\mu dq^\nu \label{DTW}
\end{align}
He showed that the metric structure
is necessary and sufficient for the hypersurfaces to be totally umbilical \cite{Yano}. The spacetime is doubly warped
if $b$ only depends on ${\bf q}$ and $a$ only depends on time $t$.\\
Ferrando, Presilla and Morales \cite{Ferrando} proved that doubly twisted spacetimes are covariantly characterized by the existence of a timelike unit, shear and vorticity free vector field: $u^i u_i=-1$ and
\begin{align}
\nabla_i u_j = \varphi (u_i u_j + g_{ij}) -u_i \dot u_j  \label{nablau}
\end{align}
where $\dot u_j=u^k\nabla_k u_j$ is the acceleration, and $\dot u_j u^j=0$.

In 1979 Bang-Yen Chen introduced twisted spacetimes, eq.\eqref{DTW} with $b=1$ \cite{Chen79}. Years later he characterized them through the existence of
a timelike torqued vector field \cite{Chen2017}: 
\begin{align}
\nabla_i \tau_j = \kappa g_{ij} +\alpha_i \tau_j, \qquad \alpha_i \tau^i=0 \label{twisted}
\end{align}
where $\kappa $ is a scalar field. 
We gave the equivalent description in terms of torse-forming time-like unit vectors, eq.\eqref{nablau} with $\dot u_i=0$, and obtained the form of the Ricci tensor \cite{ManMol2017}, and unicity of the vector, up to special cases \cite{ManTMol}.

Generalized Robertson-Walker (GRW) spacetimes were introduced in 1995 by Al\'ias, Romero and S\'anchez \cite{Alias,ManMolGRW}:
\begin{align}
ds^2 = - dt^2 +a^2(t)g^*_{\mu\nu}({\bf q})dq^\mu dq^\nu 
\end{align}
Bang-Yen Chen characterized them through the existence of a timelike concircular vector field $\nabla_i \tau_j =  \kappa g_{ij} $ \cite{Chen2014, BYChen} (the statement can be weakened, see Prop.\ref{PropW}).\\
We gave the alternative characterization \eqref{nablau} with $\dot u_i=0$ and  $\nabla_i\varphi = -u_i\dot\varphi $, and proved the useful property for the Weyl tensor \cite{ManMolJMP}: $u_m C_{jkl}{}^m =0$ if and only if $\nabla_m C_{jkl}{}^m =0$. If the Weyl tensor is zero, the spacetime is Robertson-Walker.

All these cases constitute a rich family of manifolds which are mostly studied by geometers. They also appear in physics,
as inhomogeneous extensions of the Robertson-Walker metric. In fact, the Einstein equations 
with a source of imperfect fluid with shear-free and irrotational velocity, lead to doubly-twisted metrics \cite{Maiti82,Banerjee89,Deng90,Ferrando,Msomi11,Gurses}. The Stephani universes fall in this class \cite{Stephani,Krasinski83}. 
The requirement of geodesic flow specialises the metric to twisted, with interesting applications discussed by 
Coley and McManus \cite{Coley}. Doubly warped and GRW (or warped) manifolds have an ample geometric literature \cite{Ramos,Brozos,Unal,Gebarowski,BYChen}. 

In this note we present a simple characterization that includes all such spacetimes, and discuss some 
properties of doubly torqued vectors.

\section{Another characterization}
\begin{thrm}
A Lorentzian spacetime is doubly-twisted if and only if it admits a timelike vector field, which we name `doubly torqued':
\begin{align}
\nabla_i \tau_j = \kappa g_{ij} +\alpha_i \tau_j + \tau_i \beta_j \label{doublytorqued}
\end{align}
with $\alpha_i\tau^i=0 $, $\beta_i \tau^i=0$, and $n\kappa =\nabla_i\tau^i $.
\begin{proof}
We prove the equivalence of \eqref{doublytorqued} with \eqref{nablau}.

Let $N=\sqrt{-\tau^i\tau_i}$, and introduce the vector $u_i = \tau_i /N$. Evaluate:
$\nabla_i N^2 = -2 \tau^j \nabla_i \tau_j = -2 \kappa \tau_i+ 2 \alpha_i N^2 $. Then: $\nabla_i N = - \kappa u_i +\alpha_i N$.
Next:
\begin{align*}
N\nabla_i u_j = \nabla_i \tau_j - u_j \nabla_i N
= \kappa g_{ij} + N\alpha_i u_j + N u_i \beta_j +\kappa u_iu_j - N \alpha_i u_j
\end{align*} 
Therefore: $\nabla_i u_j = (\kappa /N) (u_iu_j + g_{ij}) +u_i \beta_j $. Contraction with $u^i$ shows that $\beta_j =-\dot u_j$,
and eq.\eqref{nablau} is obtained.

Given \eqref{nablau}, the corresponding metric is \eqref{DTW}. Define $\beta_j =-\dot u_j$ and $\tau_i =  Su_i $,
where $S$ is to be found. Multiply eq.\eqref{nablau} by $S$:
$$\nabla_i (Su_j) - u_j \nabla_i S=  \varphi S(u_iu_j + g_{ij} ) + S u_i\beta_j $$
This is eq.\eqref{doublytorqued} with the vector $ \alpha_i= (\nabla_i S)/S + \varphi  u_i $. \\
We must impose the condition: 
$0= \alpha_i\tau^i = u^i\nabla_i S -  S\varphi $ i.e. $\varphi = u^i\nabla_i \log S$. \\
In the frame \eqref{DTW} it is $u_0=-b$, $u_\mu =0$, and the condition becomes $\varphi =(\partial_t S)/(Sb)$. With the Christoffel symbols in appendix we obtain $\dot u_0=u^0 (\partial_t -\Gamma_{00}^0)u_0 =0$ and $\dot u_\mu =-u^0\Gamma_{0\mu}^0 u_0 = b_\mu/b$.  The time component of \eqref{nablau},  $\nabla_0 u_0 =\varphi (u_0^2+g_{00})$,  gives $\varphi = (\partial_t a)/(ab)$. Therefore $S=a$ up to a constant factor.
\end{proof}
\end{thrm}

There is some freedom in the choice of the doubly torqued vector:
multiplication of eq.\eqref{doublytorqued} by a function $\lambda $ gives an equation for a vector $\lambda \tau_i$ that is 
orthogonal to the hypersurfaces and $\nabla_i (\lambda \tau_j) = (\lambda \kappa) g_{ij} +  (\alpha_i +\partial_i \lambda/\lambda) (\lambda \tau_j)+ (\lambda \tau_i)\beta_j$. It is doubly torqued provided that:
\begin{align}
\tau^i\partial_i \lambda =0  \label{gauge}
\end{align}

We show the relation of the special vectors $\alpha_i$ and $\beta_i$ with the scale functions $a$ and $b$ of the
metric \eqref{DTW}. \\ 
In the coordinate frame $(t,{\bf q})$, the vectors are $\tau_i=(\tau_0, 0)$, $\alpha_i=(0,\alpha_\mu)$ and 
$\beta_i=(0,\beta_\mu)$. The equations \eqref{doublytorqued} are:
$\partial_0\tau_0 -\Gamma_{00}^0\tau_0 = -\kappa b^2 $, $\partial_\mu \tau_0 -\Gamma_{\mu 0}^0\tau_0 = \tau_0 \alpha_\mu $ 
, $-\Gamma_{0\mu}^0 \tau_0 = \beta_\mu \tau_0$ and $-\Gamma_{\mu\nu}^0 \tau_0 = \kappa a^2 g^*_{\mu\nu} $.  
They can be rewritten as follows:
\begin{align}
&\beta_\mu = -\partial_\mu \log b  \label{E1}\\
&\partial_t \tau_0 = \tau_0 \partial_t \log (ab) \label{E2} \\
&\partial_\mu \tau_0 = \tau_0 (\alpha_\mu -\beta_\mu)  \label{E3}\\
&\kappa b= -\partial_t (\tau_0/b)  \label{E4}
\end{align}
The second equation is integrated: $\tau_0(t,{\bf q}) = c({\bf q}) (ab)(t,{\bf q})$, where $c$ is an arbitrary function. Then,
the first and third equation give $\alpha_\mu = \partial_\mu \log (ca)$.\\
In coordinates $(t,{\bf q})$ the condition \eqref{gauge} is $\tau^0\partial_t \lambda =0$ and implies that 
 $\lambda $ does not depend on time $t$. The transformation $(\tau_0,\alpha_\mu,\beta_\mu,\kappa)\to (\lambda\tau_0,
\alpha_\mu +\partial_\mu\lambda/\lambda, \beta_\mu, \lambda\kappa)$ leaves the above equations unchanged.
This freedom is used to put $c({\bf q})=\pm 1$. Then:
$\tau_0 (t,{\bf q}) = -(ab)(t,{\bf q})$ (if $\tau^0>0$). The other equations give: 
\begin{align}
\alpha_\mu = \frac{\partial}{\partial q^\mu} \log a, \qquad \beta_\mu=-\frac{\partial}{\partial q^\mu} \log b,  \qquad \kappa = \frac{1}{b}\frac{\partial a}{\partial t} \label{E5}
\end{align}
and establish a simple relation of the vectors with the metric. An interesting invariant is: 
\begin{align}
\tau^k\tau_k =-\frac{1}{b^2}\tau_0^2 = - a^2
\end{align}

For a doubly warped $1+n$ metric, the scale functions $b$ does not depend on time 
and $a$ does not depend on ${\bf q}$. In this case, the analysis shows that $\alpha_i $ is either zero or a gradient
orthogonal to $\tau $, that can be absorbed by a rescaling of $\tau$.\\
We obtain the characterization:
\begin{thrm}\label{T2}
A $1+n$ spacetime is doubly warped if and only if there is a timelike vector such that
$\nabla_i\tau_j = \kappa g_{ij} +\tau_i\beta_j$ with $\tau^i\beta_i=0$, and $\beta_i$ is closed.
\begin{proof}
In a doubly warped spacetime, $b({\bf q})$ and $a(t)$ are given and specialize eq.\eqref{doublytorqued}. Eq.\eqref{E5} gives: $\alpha_\mu=0$ i.e. $\alpha_i=0$. Eq.\eqref{E1} with $\partial_t b=0$ implies that $\beta_i $ is closed.

If $\alpha_i =0$ in \eqref{doublytorqued}, then eq.\eqref{E3},
$0=\partial_\mu \tau_0/\tau_0 -\partial_\mu\log b$, has solution $\tau_0 (t,{\bf q}) = F(t) b(t,{\bf q})$. The result in \eqref{E2}
gives: $\partial_t \log a = \partial_t F/F $, so that $a=a(t)$. $\beta_i $ closed becomes $\beta_\mu =-\partial_\mu r({\bf q})$. Eq.\eqref{E1} gives $b(t,{\bf q}) = \exp[r({\bf q}) + s(t)]$. The metric $ ds^2 = - e^{2r({\bf q})}e^{2s(t)}dt^2 + a^2(t) g^*_{\mu\nu} dq^\mu dq^\nu $
is doubly warped with a redefinition of time.
\end{proof}
\end{thrm}
In twisted 1+n spacetimes, $a$ depends on $t$ and ${\bf q}$, and $b=1$. Then $\beta_i=0$ and 
we recover Chen's result \eqref{twisted}. In a GRW spacetime the scale function $a$ only depends on $t$, so that also the vector 
$\alpha_i$ is zero.

\begin{prop}\label{PropW}
In a doubly twisted spacetime with doubly-torqued vector $\tau_i$,
if $\beta_i = \nabla_i \theta $ and $\tau^i\nabla_i\theta=0$ then the spacetime is conformally equivalent to a twisted spacetime.
\begin{proof}
Consider the conformal map $\hat g_{ij} = e^{2\theta} g_{ij}$. The new Christoffel simbols are 
$$\hat \Gamma_{ij}^k =
\Gamma_{ij}^k +\delta_i^k \partial_j\theta + \delta_j^k \partial_i\theta - g_{ij}g^{kl}\partial_l\theta $$ 
The vector $\hat \tau_i =
e^\theta \tau_i$ solves:
\begin{align*}
\hat \nabla_i \hat \tau_j =& \nabla_i (e^\theta \tau_j) - \hat \tau_i \partial_j\theta - \hat\tau_j \partial_i\theta + g_{ij}g^{kl}
\hat\tau_k\partial_l\theta\\
=&\hat \tau_j \partial_i\theta +e^\theta (\kappa g_{ij}+\alpha_i\tau_j +\tau_i\partial_j \theta) 
- \hat \tau_i \partial_j\theta - \hat\tau_j \partial_i\theta + \hat g_{ij}\hat g^{kl}\hat\tau_k\beta_l\\
=&(e^{-\theta} \kappa ) \hat g_{ij}+\alpha_i\hat\tau_j  
\end{align*}
The absence of $\beta_i $ characterizes a twisted spacetime.
\end{proof}
\end{prop}

A consequence of the proof is the following statement (obvious if regarded on the side of the metric): 

\begin{prop}
Conformal transformations $\hat g_{ij}=e^{2\theta} g_{ij}$ map doubly twisted  to doubly twisted spacetimes.\\
The same conformal transformation, with the condition $\tau^i\nabla_i\theta=0$, maps doubly warped to doubly warped spacetimes, or twisted to twisted spacetimes.
\begin{proof}
Given the vector $\tau_i$, consider the vector $\hat \tau_i = e^{2\theta}\tau_i$ in the new metric. It solves the equation
of a doubly torqued vector:
$$\hat \nabla_i \hat \tau_j = (\kappa + \hat \tau^k\partial_k \theta) \hat g_{ij} + (\alpha_i +\hat h_i{}^k\partial_k \theta)\hat \tau_j +\hat \tau_i  (\beta_j -\hat h_j{}^k\partial_k \theta) $$
where $\hat h_{ij}=\hat g_{ij} - \hat \tau_i\hat \tau_j/\hat \tau^2$ is a projection.

If instead the vector $\hat \tau_i = e^{\theta}\tau_i$ is considered, in the new metric it solves
$$\hat \nabla_i \hat \tau_j = (e^{-\theta} \kappa + \hat \tau^k\partial_k \theta) \hat g_{ij} + \alpha_i\hat \tau_j + \tau_i  (\beta_j -\partial_j \theta) $$
If the space is doubly warped ($\alpha_i=0$ and $\beta $ closed) then, with the condition $\tau^i\partial_i\theta=0$, the vector 
$\hat \tau_i $ solves the equation for a doubly warped spacetime. The same is true for twisted spacetimes ($\beta_i =0$).
\end{proof}
\end{prop}

\section{Conclusion}
The introduction of doubly torqued vectors, defined by the equation \eqref{doublytorqued}, has the virtue of covariantly describing a class of $1+n$ spacetimes in simple manner:
doubly-twisted ($\alpha\neq 0$, $\beta\neq 0$), doubly-warped ($\alpha =0$, $\beta \neq 0$ closed), twisted ($\alpha \neq 0$, $\beta=0$), generalized Robertson-Walker ($\alpha=\beta=0$). 

We provide some examples from the physics literature. \\
Stephani universes \cite{Krasinski83}  are conformally flat solutions of the Einstein field equations with a perfect fluid source. As in the Robertson-Walker space-time, the hypersurfaces orthogonal to the matter world-lines have constant curvature, but now its value $k$, and even its sign, changes from one hypersurface to another. The line element is doubly twisted:
$$ ds^2 = -D(t,{\bf x}) dt^2 + \frac{R^2(t)}{V^2(t,{\bf x})} (dx^2+dy^2+dz^2 ) $$
with $V=1+\frac{1}{4}\|{\bf x}-{\bf x_0}(t)\|^2$, $D=F(t)[\dot V/V - \dot R/R]$, $k=[C^2(t)-1/F^2(t)]R^2(t)$, arbitrary functions 
of time $C$, $F$, $R$ and ${\bf x_0}$.\\
Another example is the solution by Banerjee et al. \cite{Banerjee89} for a matter field with shear and vorticity free velocity and heat transfer, in a conformally flat metric:
$$ ds^2 = -V^2(t,{\bf x}) dt^2 + \frac{1}{U^2(t,{\bf x})} (dx^2 + dy^2 + dz^2)   $$
with $UV= A(t)\|{\bf x}\|^2 + {\bf A}(t){\cdot \bf x} + A_4(t)$, $U=B(t)\|{\bf x}\|^2 + {\bf B}(t){\cdot\bf x}+B_4t$, where $A$, $B$, $A_i$ and $B_i$ are arbitrary functions of time.\\
Coley studied the case with no acceleration, $V=1$, which makes the above metric twisted (\cite{Coley} eq.3.12). The same paper contains an example of GRW spacetime (eq. 1.2), which is also Bianchi VI$_0$: 
$$ds^2 = -dt^2 + X(t)^2 (dx^2 + e^{-2x}dy^2 + e^{2x}dz^2) $$ 
For a discussion and examples of doubly warped metrics, see \cite{Ramos}.

\section*{Appendix}
\noindent
These are the Christoffel symbols for the doubly-twisted metric \eqref{DTW}:
\begin{gather*}
\Gamma_{00}^0=\frac{\partial_t b}{b}, \quad 
\Gamma_{\mu,0}^0=\frac{b_\mu }{b},\quad  \Gamma_{0,0}^\mu= \frac{bb^\mu}{a^2},\quad \Gamma^\rho_{\mu,0} = 
\frac{\partial_t a}{a} \delta^\rho_\mu, \quad 
\Gamma^0_{\mu,\nu} = \frac{a\partial_t a}{b^2} g^*_{\mu\nu}, \\
\Gamma^\rho_{\mu,\nu} = \Gamma^{*\rho}_{\mu,\nu} + \frac{a_\nu}{a}  \delta^\rho_\mu + \frac{a_\mu}{a}  \delta^\rho_\nu - \frac{a^\rho}{a}   g^*_{\mu\nu} 
\end{gather*}
where $a_\mu = \partial_\mu a$ and $a^\mu = g^{*\mu\nu} a_\nu $, and the same is for $b$.

\end{document}